\documentclass[11pt]{article}
\usepackage{amsfonts, latexsym, amssymb}
\makeatletter \oddsidemargin 15mm \textwidth 148mm \textheight
9.0in \topmargin -0.50in

\def\bee{\begin{equation}}
\def\eee{\end{equation}}

\def\ti{\tilde}

\def \la{\lambda}
\def \si{\sigma}

\def \operatorname#1{\mathop{\rm #1}}

\def \0{{\theta_0}}
\def \1{{\theta_1}}

\def \op{\operatorname}

\@addtoreset{equation}{section}

\def\bm{\left(\begin{array}{cc}}
\def\em{\end{array}\right)}
\def\Bbb{\mathbb}

\newtheorem{remark}{Remark}

\def\text#1{\mbox{#1}}

\def\lim{\mathop{\mbox{lim}}\limits}

\makeatother
\begin{document}

\title{Transformations ${RS}_4^2(3)$ of the Ranks $\leq4$
and Algebraic Solutions of the
Sixth Painlev\'e Equation}
\author{F.~V.~Andreev and A.~V.~Kitaev
\thanks{E-mail: andreev@math.ksu.edu;  kitaev@pdmi.ras.ru}\\
Steklov Mathematical Institute, Fontanka 27, St.Petersburg,
191011, Russia*\\
Kansas State University, Department of Mathematics,\\
Manhattan, KS 66506, USA}

\date{July 31, 2001}
\maketitle

\begin{abstract}
Compositions of rational transformations of independent variables
of linear matrix ordinary differential equations (ODEs) with the
Schlesinger transformations ($RS$-trans\-for\-mations) are used to
construct algebraic solutions of the sixth Painlev\'e equation.
$RS$-Transformations of the ranks $3$ and $4$ of $2\times2$ matrix
Fuchsian ODEs with $3$ singular points into analogous ODE with $4$
singular points are classified.
\vspace{24pt}\vspace{24pt}\\
Short title: $RS$-transformations
\end{abstract}
\maketitle
\newpage

\setcounter{page}2
\section{Introduction}
 \label{Introd}
A considerable attention has been paid to the study of algebraic
solutions of the sixth Painlev\'e equation \cite{H1,H2,D,DM,Ma}.
Recently, in this connection, a general method for construction of
the so-called special functions of the isomonodromy type
(SFITs)\cite{K1}, which algebraically depend on one of their
variables, was suggested in \cite{K2}. SFITs include, in
particular, functions of the hypergeometric and Painlev\'e type.
With each SFIT there is an associated matrix linear ODE such that
the SFIT defines isomonodromy deformation of this linear ODE. A
key point of the proposed method is a construction of the
so-called $RS^n_m(k)$ - transformations. These transformations map
fundamental solution of $n\times n$ matrix linear ODEs with
$k$-singular points into analogous ODEs with $m$ singular points.
Each $RS$-transformation is a composition of a rational
transformation of the argument of a given linear ODE with an
appropriate Schlesinger transformation. These transformations
preserve isomonodromic property, therefore they generate
transformations of the corresponding SFITs. An important parameter
of the $RS$-transformations is their rank, $r$, which, by
definition, equals to the order of the corresponding rational
transformation, i.e., the number of its preimages counted with
their multiplicities.

In this paper we classify $RS^2_4(3)$ transformations of the ranks
$3$ and $4$ for the Fuchsian ODEs. The number of intrinsic
parameters of these transformations define whether they generate:
(1) explicit mappings of very simple SFITs, namely constants, to
the solutions of the sixth Painlev\'e equation ($P_6$); or (2)
special sets of numbers which can be interpreted as initial data
of particular solutions to $P_6$ given at some special points of
the complex plane. In fact, the solutions of the first type are
nothing but algebraic solutions to $P_6$, whilst the second ones
are, in general, transcendental solutions whose initial data at
some particular points of the complex plane can be calculated
explicitly in terms of the monodromy data of the associated linear
matrix Fuchsian ODE. Although solutions of the second type are, in
general, transcendental (non-classical in the sense of Umemura
\cite{U}), they are "less transcendental" than the other
non-classical solutions of $P_6$: due to the above mentioned
relation of the initial and monodromy data one can obtain
asymptotic expansions of these solutions in the neighborhood of
the singular points of $P_6$, $0$, $1$, and $\infty$, in terms of
their initial data provided the latter are given at the special
points. Actually, an example of solutions of the second type was
recently discussed by Manin \cite{M} in connection with some
special construction of the Frobenius manifold. The reader can
find some further details about this solution in Subsection
\ref{sub:321}.

Consider the following Fuchsian ODE with three singular points,
$0$, $1$, and $\infty$:
\begin{equation}
 \label{eq:3}
\frac{d\Phi}{d\mu}=\left( \frac{\ti A_0}\mu +\frac{\ti
A_1}{\mu-1}\right)\Phi,
\end{equation}
where $\tilde A_0$ and $\tilde A_1$ are $2\times2$ matrices
independent of $\mu$. Let us denote $\det (A_p)
=-\frac{\eta_p^2}4$, $p=0,1$. We suppose that Eq.~(\ref{eq:3}) is
normalized as follows: $\mathrm{tr}\tilde A_p=0$, $\ti A_0 +\ti
A_1 =-\frac{\eta_\infty}2\sigma_3$ where $\si_3$ is the Pauli
matrix: $\op{diag}\{1,-1\}$. Given parameters $\eta_q$, $q=0, 1,
\infty$ matrices $\ti A_q$ are fixed up to a diagonal gauge,
$c^{\sigma_3} A_pc^{-\sigma_3}$, $c\in\mathbb{C}\setminus0$. It is
well known, that solutions of Eq.~(\ref{eq:3}) can be written
explicitly in terms of the Gau\ss{} hypergeometric function (see,
e.g. \cite{J}), however, we don't use this representation here. In
this paper we consider $RS$-transformations of Eq.~(\ref{eq:3})
into the following $2\times2$ matrix ODE with four Fuchsian
singular points,
\begin{equation}
 \label{eq:4}
\frac{d\Psi}{d\lambda}=\left( \frac{A_0}\la
+\frac{A_1}{\la-1}+\frac{A_t}{\la-t}\right) \Psi,
\end{equation}
where matrices $A_l,\, l=0,1,t$ are independent of $\lambda$.
Moreover, we assume the following normalization of
Eq.~(\ref{eq:4}), $\mathrm{tr}A_l=0$,
$A_0+A_1+A_t=-\frac{\theta_\infty}2\sigma_3$. We also denote $\det
A_l=-\frac{\theta_l^2}2$. It is now well known, \cite{JM}, that
isomonodromy deformations of Eq.~(\ref{eq:4}) in general situation
(for details see \cite{K3}) define solutions of the sixth
Painlev\'e equation, $P_6$,
 \begin{eqnarray}
  \label{eq:P6}
  &\frac{d^2y}{dt^2}=\frac 12\left(\frac 1y+\frac 1{y-1}+\frac 1{y-t}\right)
\left(\frac{dy}{dt}\right)^2-\left(\frac 1t+\frac 1{t-1}+\frac 1{y-t}\right)
\frac{dy}{dt}+&\nonumber\\
&\frac{y(y-1)(y-t)}{t^2(t-1)^2}\left(\alpha_6+\beta_6\frac t{y^2}+
\gamma_6\frac{t-1}{(y-1)^2}+\delta_6\frac{t(t-1)}{(y-t)^2}\right),&
\end{eqnarray}
where $\alpha_6,\,\beta_6,\,\gamma_6,\,\delta_6\in\Bbb C$ are
parameters. We recall the relation between isomonodromy
deformations of Eq.~(\ref{eq:4}) and Eq.~(\ref{eq:P6}) following
the work \cite{JM}. Denote by $A_k^{ij}$ the corresponding matrix
elements of $A_k$. Note, that due to the normalization
$$
A_0^{12}+A_1^{12}+A_t^{12}=A_0^{21}+A_1^{21}+A_t^{21}=0,
$$
therefore equations,
$$
\frac{A_0^{ik}}{y_{ik}}+\frac{A_1^{ik}}{y_{ik}-1}+\frac{A_t^{ik}}{y_{ik}-t}=0,
$$
for $\{ik\}=\{12\}$ and $\{ik\}=\{21\}$ have in general situation
($A_1^{ik}+tA_t^{ik}\neq0$) unique solutions $y_{ik}$. These functions
solve Eq.~(\ref{eq:P6}) with the following values of the parameters,
\begin{eqnarray}
 \label{eq:12}
y_{12}(t):&&\alpha_6=\frac{(\theta_\infty-1)^2}2,\;\;
\beta_6=-\frac{\theta_0^2}2,
\;\;\gamma_6=\frac{\theta_1^2}2,\;\;\delta_6=\frac{1-\theta_t^2}2,\\
y_{21}(t):&&\alpha_6=\frac{(\theta_\infty+1)^2}2,\;\;
\beta_6=-\frac{\theta_0^2}2,
\;\;\gamma_6=\frac{\theta_1^2}2,\;\;\delta_6=\frac{1-\theta_t^2}2.
\label{eq:21}
\end{eqnarray}
One can associate with isomonodromy deformations of
Eq.~(\ref{eq:4}) one more function, the so-called $\tau$-function
\cite{O}, which plays a very important role in applications. This
function \cite{JM} is defined via the function $\sigma$,
\begin{eqnarray*}
&&\sigma(t)=\mathrm{tr}\big(((t-1)A_0+tA_1)A_t\big)+t\kappa_1\kappa_2-
\frac12(\kappa_3\kappa_4+\kappa_1\kappa_2),
\end{eqnarray*}
where
$$
\kappa_1=\frac{\theta_t+\theta_\infty}2,\;\kappa_2=
\frac{\theta_t-\theta_\infty}2,\;
\kappa_3=-\frac{\theta_1+\theta_0}2,\;\kappa_4=\frac{\theta_1-\theta_0}2.
$$
The function $\sigma$ solves the following ODE,
$$
t^2(t-1)^2{\sigma''}^2\sigma'+\left(2\sigma'(t\sigma'-\sigma)-
{\sigma'}^2-\kappa_1\kappa_2
\kappa_3\kappa_4\right)^2=(\sigma'+\kappa_1^2)(\sigma'+\kappa_2^2)
(\sigma'+\kappa_3^2)(\sigma'+\kappa_4^2),
$$
where the prime is differentiation with respect to $t$. The
$\tau$-function is defined (up to a multiplicative constant) as
the general solution of the following ODE,
$$
t(t-1)\frac{d}{dt}\ln\tau=\sigma(t).
$$

We explain now a general idea of how to construct
$RS$-transformations of Eq.~(\ref{eq:3}) to Eq.~(\ref{eq:4}).
Consider a rational transformation of the argument,
\begin{equation}
 \label{eq:rational}
\mu \equiv\mu(\lambda)=P(\lambda)/Q(\lambda),
\end{equation}
where $P(\lambda)$ and $Q(\lambda)$ are mutually prime
polynomials. The function $\mu(\lambda)$ maps Eq.~(\ref{eq:3})
with $k=3$ singular points, $0,\,1,\,\infty$, into an intermediate
Fuchsian ODE (on $\lambda$) with $3r$ singular points, where $r$
is the rank of the $RS$-transformation,
$r=\mathrm{max}(\deg\{P\},\deg\{Q\})$. If some of the parameters,
$\eta_p$, $p=0,\,1,\,\infty$, are rational numbers, say,
$\eta_0=n_0/m_0$ with the mutually prime natural numbers $n_0$ and
$m_0\geq2$, and mapping (\ref{eq:rational}) is chosen such that
some preimage of $0$ has the multiplicity proportional to $m_0$,
then this preimage is a Fuchsian singular point of the
intermediate ODE with the monodromy matrix proportional to $\pm
I$. Therefore, the latter singular point is removable by a proper
Schlesinger transformation (an apparent singularity). Thus, the
idea is to choose the parameters $\eta_p$ of Eq.~(\ref{eq:3}) and
rational mapping (\ref{eq:rational}) such that $3r-4$ points of
the intermediate equation can be removed so that one finally
arrives to Eq.~(\ref{eq:4}).

We classify $RS$-transformations up to fractional-linear
transformations of $\mu$ permutating the points $0,\,1$, and
$\infty$ and also up to fractional-linear transformations of the
variable $\lambda$ defining transformations of the set of singular
points of Eq.~(\ref{eq:4}), $0,\,1,\,\infty$, and $t$, into
$0,\,1,\,\infty$, and $\tilde t$. Clearly, $\tilde t$ is one of
the points of the orbit,
$t,\,1/t,\,1-t,\,1-1/t,\,1/(1-t),\,t/(1-t)$. In terms of the
algebraic solutions of the sixth Painlev\'e equation
fractional-linear transformations of $\mu$ are nothing but
reparametrizations of these solutions, whilst fractional-linear
transformations of $\lambda$ generally define superpositions of
so-called B\"acklund transformations of the solutions to
Eq.~(\ref{eq:P6}) with corresponding fractional-linear
transformations of $y$ and $t$. As soon as some
$RS$-transformation is constructed one can construct infinite
number of such transformations which differ only by a finite
number of Schlesinger transformations preserving singular points:
$0$, $1$, $t$, and $\infty$, of Eq.~(\ref{eq:4}). We call such
$RS$-transformations equivalent and construct, in most cases, only
one transformation representing the whole class.

For a classification of the $RS$-transformations of the rank $r$
consider partitions of $r$. Multiplicities of preimages of
$0,\,1,\,\infty$ of the rational mapping (\ref{eq:rational}) is a
triple of partitions of $r$. M\"obious invariance in $\mu$ means
that we don't distinguish triples which differ only by the
ordering of the partitions. All in all there are
$\frac16N_r(N_r+1)(N_r+2)$ of such triples, where $N_r$ is the
total number of partitions of $r$ (a number of the corresponding
Young tableaux). Our method of classification of the
$RS$-transformations can be regarded as a selection of the proper
triples of the Young tableaux. It consists of three steps:
\begin{enumerate}
\item
A sieve-like procedure with the goal to get rid of the triples
which generate more than $m=4$ non-removable singularities of the
intermediate ODE. To calculate the number of triples which pass
through the sieve let us introduce some notation. For each
partition ${\cal P}$ denote ${\cal M}$ a maximal subset with the
greatest common divider greater than $1$. In the case when there
are several such subsets, ${\cal M}$ is anyone of them. Denote
$a_j$ a number of the Young tableaux with $\mathrm{card}({\cal
P\setminus M})=j$. Clearly, the sieve is passed by those triples
which satisfy the condition, $j_1+j_2+j_3\leq4$, where $j_k$ means
a value of the parameter $j$ for the $k$-th partition of the
triple. The total number of the triples which pass through the
sieve is
\begin{equation}
 \label{eq:sieve}
\begin{array}{l}
\frac{a_0(a_0+1)}2\left(\frac{a_0+2}3+a_1+a_2+a_3+a_4\right)+
\frac{a_1(a_1+1)}2\left(a_0+\frac{a_1+2}3+a_2\right)\\
+\frac{a_2(a_2+1)}2a_0
+a_0a_1(a_2+a_3)-1;
\end{array}
\end{equation}
The last term in Eq.~(\ref{eq:sieve}), $-1$, is related with the
fact that equation
$$
x^r+y^r=z^r
$$
has no solutions in mutually prime polynomials, $x$, $y$, and $z$,
for $r\geq2$.
\item
The aim of this stage is to choose among the triples selected at
the first step those ones for which there exist corresponding
rational mappings (\ref{eq:rational}). Denote
$i=\mathrm{card}{\cal M}$, then Eq.~(\ref{eq:4}) with arbitrary
parameter $t$ can exist only in the case when
\begin{equation}
 \label{eq:r+3}
i_1+j_1+i_2+j_2+i_3+j_3\geq r+3,
\end{equation}
where the subscripts denote parameters $i$ and $j$ characterizing
partitions in the triple.
In the case
\begin{equation}
 \label{eq:r+2}
i_1+j_1+i_2+j_2+i_3+j_3=r+2
\end{equation}
only $RS$-transformations with a parameter $t$ equal to some special
number could exist;
\item
The final stage is the construction of the $RS$-transformation.
This stage includes an analysis of how many $RS$-transformations
can be constructed for a given partition: sometimes there are few
different $RS$ - transformations due to the ambiguity of the
choice of the set ${\cal M}$ for some partitions.
\end{enumerate}
One more question which sounds naturally in connection with the
$RS$-transformations is as follows: which transformations of the rank $r$
can be presented as a superposition of $RS$-transformations of lower ranks
$r_1$, $r_2$,...$r_N$? Clearly in the latter case one writes
$$
r=r_1r_2\cdot\ldots\cdot r_N.
$$

The classification of $RS$-transformations of the rank $2$, $3$
and $4$ is given in the following sections.
\section
{$RS$-transformations of the rank $2$}
 \label{sec:2}
We apply the scheme suggested in Introduction for $r=2$ ($N_2=2$).
After the first stage from the total number
$4=\frac16\cdot2\cdot3\cdot4$ of the triples of partitions of $r$
we are left with $2$ triples, $(2|1+1|1+1)$ and $(2|1+1|2)$, since
$a_0=a_2=1$ and $a_1=a_3=a_4=0$. At the second stage we construct
rational mappings corresponding to these triples. The rational
mapping which corresponds to the second triple clearly has a very
simple form, $\mu=\lambda^2$. The parameter $t=-1$, so that this
transformation does not generate any algebraic solution of the
sixth Painlev\'e equation. The latter fact also follows from
Eq.~(\ref{eq:r+2}), since $i_1=i_3=1$, $j_1=j_3=0$, $i_2=0$, and
$j_2=2$. However, as it is mentioned in Introduction (see also
\cite{K2}), this transformation can be interpreted as some
property of special solutions of the sixth Painlev\'e equation.

To the first triple, more precisely, to the triple $(1+1|2|1+1)$,
there corresponds the following rational mapping,
\begin{equation}
 \label{eq:P2mu}
\mu =\rho \frac{\la (\la-1)}{\la -t}
\qquad\mathrm{and}\qquad
\mu-1=\rho \frac{(\la-a)^2}{\la -t},
\end{equation}
where
$$
t=\frac{s^2(s-1)^2}{(s^2-2s-1)(s^2+1)},\quad
a=\frac{s(s-1)}{s^2+1},\quad \rho=\frac{s^2+1}{s^2-2s-1},
$$
with arbitrary $s\in\mathbb{C}$. This fact is consistent with
Eq.~(\ref{eq:r+3}), in this case $i_1=i_3=0$, $j_1=j_3=2$,
$i_2=1$, and $j_2=0$. Rational mapping (\ref{eq:P2mu}) generates
$RS$-transformation which we denote $RS_4^2(1+1|2|1+1)$. This
transformation exists for arbitrary parameters
$\eta_0,\,\eta_\infty\in\mathbb{C}$ and parameter $\eta_1=1/2$.
and reads as follows,
$$
\Psi(\la)=\left( J_\infty^* \sqrt{\la-a} +J_a ^* \frac
1{\sqrt{\la-a}}\right) \Phi(\mu),
$$
where
$$
J_\infty^*=\!\left(\!\begin{array}{cc}
0&0\\
0&1\end{array}\!\right)\!,\quad J_a^*=\!\left(\!\begin{array}{cc}
1&0\\
0&\frac{(2\eta_\infty+2\eta_0+1)(s^2-2s-1)}
{16\eta_\infty(\eta_\infty+1)(s^2+1)}
\end{array}\!\right)\!
\left(\!\begin{array}{cc}
1&\frac{2\eta_\infty-2\eta_0+1}{2\eta_\infty+2\eta_0-1}\\
2\eta_\infty+2\eta_0-1&2\eta_\infty-2\eta_0+1
\end{array}\!\right)\!.
$$
The values of the $\theta$-parameters are,
$$
\theta_0=\eta_0,\quad \theta_1=\eta_0,\quad
\theta_t=\eta_\infty,\quad \theta_\infty=\eta_\infty+1.
$$
Corresponding algebraic solutions of the sixth Painlev\'e equation
and related functions $\sigma(t)$ and $\tau(t)$ are as follows:
\begin{eqnarray*}
&\phantom{Aa}y_{12}(t)=\frac{(s-1)s}{s^2-2s-1}=t+\sqrt{t^2-t},
\phantom{AAAAAAAAAAAAAAAAAAAAAAAAAAA}&\\
&y_{21}(t)=y_{12}(t)
\frac{((2\eta_\infty+3+2\eta_0)(s^2+1)-(2+4\eta_0)(s+1))
((2\eta_\infty+3-2\eta_0)(s^2+1)-(2-4\eta_0)(s+1))}
{((2\eta_\infty+3)^2-4\eta_0^2)(s^2+1)^2-4(4\eta_0^2-1)(s^3-s)},&
\end{eqnarray*}
\begin{eqnarray*}
\sigma(t)&=&
-\frac{(4\eta_0^2+(1+2\eta_\infty)^2)(s^2+1)^2-16\eta_0^2(s^3-s)}
{16(s^2+1)(s^2-2s-1)},\\
\tau(t)&=&
(s^3-s)^{-\frac12\eta_0^2-\frac12(\frac12+\eta_\infty)^2}
(s^2-2s-1)^{(\frac12+\eta_\infty)^2}(s^2+1)^{\eta_0^2}.
\end{eqnarray*}
\begin{remark}
Hereafter we omit the multiplicative parameter $C$ in formulae for
the $\tau$-function $(\tau(t)\to C\tau(t))$.
\end{remark}
Since the function $y_{12}$ is independent of the parameters
$\eta_0$ and $\eta_\infty$ corresponding terms in
Eq.~(\ref{eq:P6}), proportional to $\eta_0^2$ and $\eta_\infty^2$
should vanish for this solution. Therefore, $y_{12}$ solves the
following algebraic equation:
$$
\frac{t}{y_{12}^2}-\frac{t-1}{(y_{12}-1)^2}=0\quad\mathrm{or,\;
equivalently,}\quad
1-\frac{t(t-1)}{(y_{12}-t)^2}=0.
$$
\section
{$RS$-transformations of the rank $3$}
 \label{sec3}
In the case $r=3$ ($N_3=3$) the total number of different triples
is $\frac16\cdot3\cdot4\cdot5=10$. According to
Eq.~(\ref{eq:sieve}) five triples survive after the first stage,
since $a_0=a_1=a_3=1$ and $a_2=a_4=0$. They are $(1+1+1|2+1|3)$,
$(1+1+1|3|3)$, $(2+1|2+1|2+1)$, $(2+1|2+1|3)$, and $(2+1|3|3)$. As
it follows from Eqs.~(\ref{eq:r+3}) and (\ref{eq:r+2}) two
triples, $(1+1+1|2+1|3)$ and $(2+1|2+1|2+1)$, define
$RS$-transformations with the arbitrary parameter $t$; two
triples, $(1+1+1|3|3)$ and $(2+1|2+1|3)$, define
$RS$-transformations with fixed $t$; and the last triple,
$(2+1|3|3)$,  does not define any $RS$-transformation.
\begin{subsection}
{$RS$-transformations with arbitrary $t$}
\end{subsection}
\subsubsection
{$RS_4^2(2+1|2+1|2+1)$}
 \label{sss:311}
$R$-Transformation reads:
$$
\mu=\frac{\rho\lambda(\lambda-t)^2}{(\lambda-b)^2}
\qquad\mathrm{and}\qquad
\mu-1=\rho\frac{(\la-1)(\la-c)^2}{(\la-b)^2},
$$
where
$$
t=1-s^2,\quad c=(s+1)^2,\quad \rho=\frac1{(2s+1)^2},\quad
b=\frac{(s+1)^2}{2s+1},
$$
with arbitrary $s\in\mathbb{C}$. We define $RS$-transformation for
arbitrary value of $\eta_0\in\mathbb{C}$, $\eta_1=1/2$, and
$\eta_\infty=-1/2$, as follows:
$$
\Psi(\la) =\left( J_c \sqrt{\frac{\la-c}{\la-b}} +J_b
\sqrt{\frac{\la-b}{\la-c}}\right) \Phi(\mu).
$$
In the previous formula,
$$
J_c=\bm 0 &\frac{\eta_0}{\eta_0-1} \\ 0 & 1\em,\quad
J_b=J_c^*.
$$
Hereafter, we use the following matrix operation ${}^*$,
$$
A=\bm a_{11} & a_{12}\\ a_{21} & a_{22} \em,
\quad A^*=
\bm a_{22} & -a_{12}\\ -a_{21} & a_{11} \em.
$$
The $\theta$-parameters read:
$$
\theta_0=\eta_0,\quad \theta_1=1/2,
\quad \theta_t=2\eta_0,\quad \theta_\infty=-1/2.
$$
Corresponding solutions of $P_6$ are
\begin{eqnarray}
y_{12}&=&
\frac{(1-s^2)(\eta_0^2(2s+1)^2-1)}
{\eta_0^2(2s+1)^3-1-2s^3},\nonumber\\
 \label{yold}
y_{21}&=&
\frac{1-s^2}{2s+1},
\end{eqnarray}
with the following associated functions
$$
\sigma= -\frac 58\eta_0^2-\eta_0^2s+\frac 1{16}s^2
\quad\mathrm{and}\quad
\tau=\frac1{(1-s^2)^{\frac 1{16}}}
\left(\frac{(1-s)^{\frac{13}8}}{(1+s)^{\frac 38}s^{\frac
54}}\right) ^{\eta_0^2}.
$$
By analogous arguments as at the end of Section \ref{sec:2} one
finds that $y_{21}$ solves the algebraic equation
$$
\frac1{y_{21}^2}+\frac{4(t-1)}{(y_{21}-t)^2}=0.
$$
\subsubsection
{$RS_4^2(1+1+1|3|2+1)$}
$R$-Transformation reads,
$$
\mu=\rho\frac{\la(\la-1)(\la-t)}{(\la-b)^2}
\qquad\mathrm{and}\qquad
\mu-1=\rho\frac{(\la-c)^3}{(\lambda-b)^2},
$$
with
$$
\rho =\frac{(1-s)(s+1)^3}{s^2+s+1}, \quad
b =\frac1{(1-s)(s^2+s+1)},\quad
c=\frac1{1-s^2},\;\;
\mathrm{and}\;\;
t=\frac{2s+1}{(1-s)(s+1)^3},
$$
where $s\in\mathbb{C}$. To define the $RS$-transformation we
choose $\eta$-parameters,
$$
\eta_1=\frac 13,\qquad \eta _\infty = -\frac 12,
$$
and arbitrary $\eta_0\in\mathbb{C}$. Then $RS$-transformation has
the following form,
$$
\Psi(\lambda)= \left(\sqrt{\frac{\la-c}{\la-b}}J_c+
\sqrt{\frac{\lambda -b}{\la-c}}J_b\right)\Phi(\mu),
$$
where
$$
J_b=\bm
1&-{\displaystyle\frac{6\,{\eta _{0}} + 1}{6\,{\eta _{0}}- 5}}\\[2ex]
0&0
\em,
\qquad J_c=J_b^*.
$$
The function $\Psi(\lambda)$ has the following parameters of formal monodromy:
$$
\theta_0=\theta_1=\theta_t=\eta_0,\quad \theta_\infty=-\frac12.
$$
We find
\begin{eqnarray*}
y_{12}\!\!&=&\!\!\frac{(2s+1)(36\eta_0^2(s^2+s+1)^2-(s^2-5s-5)^2)}
{(s+1)((36\eta_0^2-25)(s^6+3s^5+3s+1)+6(36\eta_0^2-7)(s^4+s^2)+
(252\eta_0^2+41)s^3)},\\
y_{21}\!\!&=&\!\!\frac{(2s+1)}{(s+1)(s^2+s+1)}.
\end{eqnarray*}
and
\begin{eqnarray*}
\sigma(t)&=&
\frac{(108\eta_0^2+1)(s^4+2s^3+2s+1)+6(36\eta_0^2-1)s^2}
{288(s+1)^3(s-1)},\\
\tau(t)&=&\left((s+1)\Big(\frac1s+1\Big)\right)^{\frac1{24}}
\left((s-1)\Big(\frac1s-1\Big)\right)^{\frac{\eta_0^2}{2}}
\left((s+2)\Big(\frac1s+2\Big)\right)^{-\frac1{32}+\frac58\eta_0^2}.
\end{eqnarray*}
Note, that the function $y_{21}=y_{21}(t)$ solves the following algebraic
equation,
$$
\frac{t(t-1)}{(y_{21}-t)^2}-\frac{(t-1)}{(y_{21}-1)^2}+\frac{t}{y_{21}^2}=0.
$$
\begin{remark}
In the case $\eta_0=0$ solution $y_{12}(t)$ coincide with the
so-called Tetrahedron Solution found in {\rm\cite{D,DM}}
\footnote{There is a misprint in the sign of $x$ in \cite{DM} p. 139
formula ($A_3$).}:
$$
y(t)=\frac{(h-1)^2(1+3h) (9h^2-5)^2}{(1+h)(25-207h^2 +1539 h^4 +243 h^6)},
\quad
t =-\frac{(h-1)^3 (1+3h)}{(h+1)^3 (1-3h)},
$$
where parameter $h=s/(s+2)$. Clearly, it can be obtained by a B\"acklund
transformation from a simpler solution $y_{12}(t)$ since the same is true
for solution $y_{21}(t)$ with arbitrary $\eta_0$.
\end{remark}
\begin{subsection}
{$RS$-transformations with fixed $t$}
\end{subsection}
\subsubsection
{$RS_4^2(1+1+1|3|3)$\label{sub:321}}
$R$-transformation reads:
$$
\mu=\mp3i\sqrt{3}\lambda(\lambda-1)(\lambda-t)
\qquad\mathrm{and}\qquad
\mu-1=\mp3i\sqrt{3}(\lambda-c)^3,
$$
with
$$
t=\frac12\pm i\frac{\sqrt{3}}2,\quad\mathrm{and}\quad
c=\frac12\pm i\frac{\sqrt{3}}6,
$$
where, and thereafter in this subsection, one should take in all
formulae either upper or lower signs correspondingly; so that we
actually have two $R$-transformations.

For each $R$-transformation there are two different
$RS$-transforms which correspond to the following choice of the
$\eta$-parameters: $\eta_1=1/3$ or $\eta_1=2/3$, whereas $\eta_0$
and $\eta_\infty$ are arbitrary complex numbers in both cases.
Resulting $\theta$-parameters are:
$\theta_0=\theta_1=\theta_t=\eta_0$, in both cases,  and
$\theta_\infty=3\eta_\infty+1$ or $\theta_\infty=3\eta_\infty$,
correspondingly. We haven't checked yet whether one of these,
formally different, $RS$-transformations can be obtained from the
other by simply making the shift of
$\eta_\infty\to\eta_\infty-1/3$. Consider the first
$RS$-transformation ($\eta_1=1/3$):
\begin{equation}
\Psi(\lambda)=\left(\left(\begin{array}{cc}
0&0\\
0&1
\end{array}\right)\sqrt{\la-c}+ \left(\begin{array}{cc}
1&-p\\
0&0
\end{array}\right)\frac1{\sqrt{\la-c}}\right)\Phi(\mu),\qquad
p=\frac{3\eta_0-3\eta_\infty-1}{3\eta_0+3\eta_\infty-1}.
\label{PsiManin}
\end{equation}
The function $\Psi(\lambda)$ has the following parameters of
formal monodromy:
\begin{equation}
 \label{th}
\theta_0=\theta_1=\theta_t=\eta_0,\qquad
\theta_\infty=3\eta_\infty+1.
\end{equation}
The residue matrices of Eq.~(\ref{eq:4}) read:
\begin{eqnarray*}
A_0=\left(\!\!\begin{array}{cc}
-\frac12\eta_\infty-\frac16&\frac{6\eta_\infty p}{3\pm i\sqrt{3}}\\
\frac{(3\pm
i\sqrt{3})q}{216\eta_\infty}&\frac12\eta_\infty+\frac16
\end{array}\!\!\right),&&
A_1=\left(\!\!\begin{array}{cc}
-\frac12\eta_\infty-\frac16&-\frac{6\eta_\infty p}{3\mp i\sqrt{3}}\\
-\frac{(3\mp
i\sqrt{3})q}{216\eta_\infty}&\frac12\eta_\infty+\frac16
\end{array}\!\!\right),\\
A_t=\left(\!\!\begin{array}{cc}
-\frac12\eta_\infty-\frac16&\pm i\eta_\infty\sqrt{3}p\\
\frac{\mp i\sqrt{3}q}{108\eta_\infty}&\frac12\eta_\infty+\frac16
\end{array}\!\!\right),&&q=9(\eta_0+\eta_\infty)^2-1.
\end{eqnarray*}
Using these formulae we find:
\begin{eqnarray*}
y_{12}(\frac12\pm i\frac{\sqrt{3}}2)\!\!&=&\!\!\frac12\pm
i\frac{\sqrt{3}}6, \qquad y'_{12}(\frac12\pm
i\frac{\sqrt{3}}2)=\frac13,\\
y_{21}(\frac12\pm i\frac{\sqrt{3}}2)\!\!&=&\!\!\infty,\quad
q\neq0.
\end{eqnarray*}
In the case $q=0$ the value of $y_{21}(1/2\pm i\sqrt{3}/2)$ can't
be determined.

It is worth to notice that making a special choice
of the $\eta$-parameters in this $RS$-transformation
$\eta_0=0$, $\eta_1=1/3$, and $\eta_\infty=1$, we arrive to the
solution considered by Manin (see p.~81 of {\rm\cite{M}}).

Indeed, in \cite{M}, a Frobenius manifold of dimension three is described
by a solution of P6 with the initial data
$$
y(t)=\frac12+i\frac{\sqrt 3}6
\qquad\mathrm{and}\qquad
y'(t)=\frac13
$$
given at the point $t=\frac 12+\frac{\imath \sqrt 3}2$ with the parameters
$$
(\alpha,\beta,\gamma,\delta)=(\frac 92,0,0,\frac 12).
$$
For $\theta$-parameters it means $\theta_0=\theta_1=\theta_t=0$
and $\theta_\infty$ is either $-2$ or $4$. In the following
construction $\theta_\infty=4$. We obtain these
$\theta$-parameters by putting $\eta_0=0$, $\eta_1=1/3$, and
$\eta_\infty=1$ in Eq.~(\ref{th}).

Since, our construction allows us to find $\Psi$-function {\it explicitly}, see (\ref{PsiManin}),
we can write down the monodromy data. They are:
\begin{eqnarray*}
&M_0= \bm 1+i\sqrt 3 & -\frac{27}{8\pi^3}\Gamma(\frac
23)^6(\sqrt3+i)
\\
-\frac 29\frac{\pi^3(-i+\sqrt 3)}{\Gamma (\frac 23)^6}
&1-i \sqrt 3\em,&\\
& M_1 = \bm 1+7i\sqrt3 & \frac{27}{8\pi^3}\Gamma(\frac 23)^6
(3\sqrt 3-13i)\\
\frac 29\frac{\pi^3(3\sqrt 3+13i)}
{\Gamma(\frac 23)^6}& 1-7i\sqrt3 \em,&\\
&M_t = \bm 1+i \sqrt 3& \frac{27}{8\pi^3}
\Gamma (2/3)^6(-i +\sqrt3 )\\
\frac 29\frac{\pi^3(\sqrt3+i)}{
\Gamma (\frac 23)^6}& 1-i\sqrt3\em,&\\
&M_\infty = \bm 1+3i\sqrt 3& -\frac {81}{4\pi^3}
i\Gamma (\frac 23)^6\\
\frac 43\frac{i\pi^3}{\Gamma(\frac 23)^6}& 1-3i\sqrt3 \em.&
\end{eqnarray*}
These matrices $M$ coincide with the corresponding monodromy
matrices for the representation of the fundamental group defined
in \cite{J}. They satisfy the following cyclic relation: $M_\infty
M_1 M_t M_0=1$. Note, that since $\theta_\infty$ is an integer
number, $M_\infty$ is not a diagonalizable  matrix.

\subsubsection
{$RS_4^2(2+1|3|2+1)$}
Corresponding $R$-transformation reads,
$$
\mu= -\frac{\lambda(\lambda-1)^2}{3(\lambda-1/9)^2}
\qquad\mathrm{and}\qquad
\mu-1=-\frac{(\lambda+1/3)^3}{3(\lambda-1/9)^2}.
$$
Using this $R$-transformation one can define three different
$RS$-transformations corresponding to the following following
choice of the $\eta$-parameters:
\begin{enumerate}
\item
$\eta_0$ and $\eta_1$ are arbitrary, $\eta_\infty=\frac 12$. The
$\theta$-parameters in Eq.~(\ref{eq:4}) with $t=-1/3$ are as
follows: $\theta_0=\eta_0$, $\theta_1=2\eta_0$,
$\theta_t=3\eta_1$, and $\theta_\infty=-\frac12$;
\item
$\eta_0$ and $\eta_\infty$ are arbitrary, $\eta_1=\frac13$. The
$\theta$-parameters in Eq.~(\ref{eq:4}) with $t=1/9$ are as
follows: $\theta_0=\eta_0$, $\theta_1=2\eta_0$,
$\theta_t=2\eta_\infty$, and $\theta_\infty=\eta_\infty-1$;
\item
$\eta_0$ and $\eta_\infty$ are arbitrary, $\eta_1=\frac23$. The
$\theta$-parameters in Eq.~(\ref{eq:4}) with $t=1/9$ are as
follows: $\theta_0=\eta_0$, $\theta_1=2\eta_0$,
$\theta_t=2\eta_\infty$, and $\theta_\infty=\eta_\infty$.
\end{enumerate}
\begin{section}
{$RS$-transformations of the rank $4$}
 \label{sec:4}
\end{section}
In the case $r=4$ ($N_4=5$) the total number of different triples
are $\frac16\cdot5\cdot6\cdot7=35$. According to
Eq.~(\ref{eq:sieve}) twenty triples survive after the first stage
, since $a_0=2$, $a_1=a_2=a_4=1$, and $a_3=0$. As follows from
Eqs.~(\ref{eq:r+3}) and (\ref{eq:r+2}) two triples,
$(1+1+1+1|2+2|2+2)$ and $(2+1+1|2+1+1|2+2)$, define
$RS$-transformations with arbitrary parameter $t$ which have an
additional parameter; five triples, $(1+1+1+1|2+2|4)$,
$(2+1+1|2+1+1|4)$, $(2+1+1|2+2|2+2)$, $(2+1+1|2+2|3+1)$, and
$(2+1+1|3+1|3+1)$, define $RS$-transformations with arbitrary $t$;
seven triples, $(1+1+1+1|4|4)$, $(2+1+1|2+2|4)$, $(2+1+1|3+1|4)$,
$(2+2|2+2|2+2)$, $(2+2|2+2|3+1)$, $(2+2|3+1|3+1)$, and
$(3+1|3+1|3+1)$, correspond to the $RS$-transformations with fixed
$t$; finally, the following six triples, $(2+1+1|4|4)$,
$(2+2|2+2|4)$, $(2+2|3+1|4)$, $(2+2|4|4)$, $(3+1|3+1|4)$,
$(3+1|4|4)$, do not define any $RS$-transformation.
\begin{subsection}
{$RS$-Transformations with arbitrary $t$}
\end{subsection}
\subsubsection
{$RS_4^2(3+1|3+1|2+1+1)$}
$R$-Transformation reads,
$$
\mu=\frac{\rho\lambda(\lambda-a)^3}
{(\lambda-b)^2(\lambda-1)}
\qquad\mathrm{and}\qquad
\mu-1=\frac{\rho(\la-t)(\la-c)^3}{(\la-b)^2(\la-1)},
$$
where
\begin{eqnarray*}
\rho&=&\frac{(1-2s)^3}{(1-3s^2)^2(1-3s)^2},\qquad
a=\frac{(1-3s^2)}{(1-2s)}(3s^2-2s+1),\\
b&=&\frac{(1-3s^2)}{(1-3s)}(3s^2-3s+1),\qquad
c=1-3s^2,
\end{eqnarray*}
and
$$
t=\frac{(1-3s^2)}{(1-2s)^3}(3s^2-3s+1)^2,
$$
with arbitrary $s\in\mathrm{C}$. We consider below two different
choices of the $\eta$-parameters, generating however equivalent
seed $RS$-transformations, which are associated with this
$R$-transformation. The reason, why we present here two equivalent
constructions, is explained in Remark \ref{rem:4.1}:

{\bf\thesubsubsection.A}. The first $RS$-transformation can be
defined by making the following choice of the $\eta$-parameters in
Eq.~(\ref{eq:3}):
\begin{equation}
 \label{eq:eta1}
\eta_0=1/3,\qquad\eta_1=1/3,\qquad\eta_\infty=-1/2.
\end{equation}
and reads,
\begin{eqnarray*}
\Psi(\lambda)=
\left( J_1^* \sqrt{\frac{\la-b}{\la-1}}
+J_b^* \sqrt{\frac{\la-1}{\la-b}}\right)
\left( J_a \sqrt{\frac{\la-a}{\la-c}}
+J_c \sqrt{\frac{\la-c}{\la-a}}\right)
\Phi(\mu),
\end{eqnarray*}
where
$$
J_a=\frac 12 \bm 1 & 1\\ 1 & 1 \em, \qquad J_c=J_a^*,
$$
and
$$
J_b=\frac 1{6s(1-2s)}\bm 1-3s^2 & -(1-3s)(s-1)\\
\frac{(1-3s)(1-3s^2)}{s-1} & -(1-3s)^2
\em,\qquad
J_1=1-J_b.
$$
It results in the following values of the $\theta$-parameters in
Eq.~(\ref{eq:4}):
$$
\theta_0=1/3,\quad \theta_1=1/2,
\quad \theta_t=1/3,\quad \theta_\infty=-1/2,
$$
and generates the following solutions of $P_6$ and associated functions:
\begin{eqnarray*}
y_{12}&=&-\frac{(3s^3-3s^2+3s-1) (3s^2-3s+1)(9s^5-15s^4-30s^3+60s^2-35s+7)}
{(1-3s)(1-2s)(135s^6-378s^5+441s^4-288s^3+129s^2-42s+7)},\\
y_{21}&=&\frac{(3s^2-3s+1)(3s^2-2s+1)}{(1-3s)(1-2s)},
\end{eqnarray*}
\begin{eqnarray*}
\sigma&=&\frac{432s^6-972s^5+765s^4-176s^3-81s^2+54s-9}{144(1-2s)^3},\\
\tau&=&\frac{(3s-2)^{\frac 1{16}}
(1-2s)^{\frac 1{12}}}
{s^{\frac 3{16}}(3s^2-3s+1)^{\frac 1{18}}
(1-3s^2)^{\frac {13}{144}}}.
\end{eqnarray*}

{\bf\thesubsubsection.B}.
Another choice of the $\eta$-parameters is as follows:
\begin{equation}
 \label{eq:eta2}
\eta_0=1/3,\qquad\eta_1=2/3,\qquad\eta_\infty=-1/2.
\end{equation}
However, making a proper Schlesinger transformation of
Eq.~(\ref{eq:3}) and further transformation which is related with
the reflection, $\eta_\infty\to-\eta_\infty$ we see that
corresponding $RS$-transformations and, hence, algebraic solutions
are equivalent to the ones constructed in {\bf A}. Below we
present a bit different construction of the $RS$-transformation
starting with the choice of the $\eta$-parameters given by
(\ref{eq:eta1}); this, however, results in the same algebraic
solutions as for the choice (\ref{eq:eta2}):
$$
\Psi(\lambda)=\left(J_t^*\sqrt{\frac{\la-b}{\la-t}}
+\hat J_b^*\sqrt{\frac{\la-t}{\la-b}}\right)
\left(J_a\sqrt{\frac{\la-a}{\la-c}}
+J_c\sqrt{\frac{\la-c}{\la-a}}\right)
\Phi(\mu),
$$
where $\mu$, $J_a$, and $J_c$ are the same as in {\bf A},
$$
\hat J_b=\frac 1{2s}\bm -s+1 &  -s+1\\ -1+3s & -1+3s\em, \qquad J_t=1-J_b.
$$
The function $\Psi$ solves Eq.~(\ref{eq:4}) with the following
values of the $\theta$-parameters:
$$
\theta_0=1/3,\quad \theta_1=1/2,
\quad \theta_t=2/3,\quad \theta_\infty=-1/2,
$$
and generates the following solutions of $P_6$ and related functions:
\begin{eqnarray}
y_{12}&=&
-\frac{(5s^2-4s+1)(3s^2-3s+1)(45s^4-102s^3+96s^2-42s+7)}
{(135s^5-405s^4+450s^3-240s^2+63s-7)(1-2s)^2},\nonumber\\
y_{21}&=&
\frac{(s-1)(1-3s)(1-3s^2)(3s^2-3s+1)}
{(9s^3-9s^2+3s-1)(1-2s)^2}.
 \label{ynew}
\end{eqnarray}
\begin{eqnarray*}
\sigma&=&
\frac{108s^6-216s^5+9s^4+244s^3-225s^2+90s-15}
{144(1-2s)^3},\\
\tau&=&\frac{(1-2s)^{\frac 13} (1-3s^2)^{\frac {17}{144}}}
{s^{\frac 5{16}}
(3s-2)^{\frac 1{16}}
(3s^2-3s+1)^{\frac 5{36}}}.
\end{eqnarray*}
\begin{remark}
 \label{rem:4.1}
It is interesting to note that for the $\theta$-parameters
considered in this subsection, i.e., $\theta_0=1/3$,
$\theta_1=1/2$, $\theta_\infty=-1/2$, and $\theta_t=1/3$ or
$\theta_t=2/3$, we have constructed here and in Subsection
{\rm\ref{sss:311}} two different algebraic solutions: given by
Eqs.~{\rm(\ref{yold})} and {\rm(\ref{ynew})}. Indeed, solution
{\rm(\ref{ynew})} has three finite poles at points $t_k=t(s_k)$,
where  $s_1=1/3+\sqrt[3]{2}/3$ and $s_{2,3}=1/3-\sqrt[3]{2}(1\pm
i\sqrt{3})/6$ $(t_1\approx1.0577\ldots,
t_{2,3}\approx0.8391\ldots\pm i0.3357\ldots)$, while
{\rm(\ref{yold})} has only one pole at $3/4$.
\end{remark}
\subsubsection
{$RS_4^2(2+2|2+2|2+1+1)$}
$R$-Transformation is as follows:
$$
\mu=\frac{\rho\lambda^2(\lambda-1)^2}
{(\lambda-t)(\lambda-b)^2}
\qquad\mathrm{and}\qquad
\mu-1=\frac{\rho(\la-c_1)^2(\la-c_2)^2}{(\la-b)^2(\la-t)},
$$
where
$$
\rho =\frac{2s^2-1}{4(2s^2+2s+1)},\quad
b=\frac{s(2s+1)}{2s^2-1},\quad
c_1=\frac{2s^2}{2s^2-1},\quad
c_2=\frac{(2s+1)^2}{2s^2-1},
$$
and
$$
t=\frac{s^2(2s+1)^2}{(2s^2-1)(2s^2+2s+1)},
$$
with $s\in\mathbb{C}$. For the choice:  $\eta_0\in\mathbb{C}$ is
arbitrary, $\eta_1=1/2$, and $\eta_\infty=-1/2$,
$RS$-transformation reads,
$$
\Psi(\lambda)=
\left( J_t^* \sqrt{\frac{\la-c_2}{\la-t}}
+J_{c_2}^* \sqrt{\frac{\la-t}{\la-c_2}}\right)
\left( J_b \sqrt{\frac{\la-b}{\la-c_1}}
+J_{c_1} \sqrt{\frac{\la-c_1}{\la-b}}\right)
\Phi(\mu),
$$
where
$$
J_{c_1}=J_{c_2}=\bm 0 & \frac{\eta_0}{\eta_0-1} \\
0 & 1 \em , \quad
J_b=J_{c_1}^*, \quad J_t=1-J_{c_2}.
$$
Parameters:
$$
\theta_0=2\eta_0,\quad \theta_1=2\eta_0, \quad
\theta_t=1/2,\quad \theta_\infty=-1/2.
$$
Solutions of $P_6$ and related functions $\sigma$ and $\tau$ are as follows:
\begin{eqnarray*}
y_{12}&=&
\frac{
s(2s+1)(4\eta_0^2(2s^2+2s+1)^2-s^2(2s+1)^2)}
{(2s^2+2s+1)((4\eta_0^2-1)(2s^2+2s+1)^2+3s(s+1)(2s+1))},\\
y_{21}&=&
\frac{s(2s+1)}{2s^2+2s+1}.
\end{eqnarray*}
\begin{eqnarray*}
\sigma&=&
-\frac{(2s^2+2s+1)\eta_0^2}{2s^2-1},\\
\tau&=&\left(\frac{(2s^2-1)^2}{s(s+1)(2s+1)}\right)^{2\eta_0^2}.
\end{eqnarray*}
\begin{remark}
The function $y_{21}$ solves the following algebraic equation,
$$
\frac{t}{y_{21}^2}-\frac{(t-1)}{(y_{21}-1)^2}=0,
$$
and therefore, can be written as $y_{21}=t-\sqrt{t^2-t}$. Thus the
solutions constructed in this subsection are not new, they
coincide with the ones obtained in Section {\rm\ref{sec:2}} in the
case $\eta_\infty=-1/2$ and $\eta_0\to2\eta_0$. The explanation of
this fact is that corresponding $RS$-transformation is actually a
combination of a quadratic transformation for
Eq.~{\rm(\ref{eq:3})} with the quadratic transformation obtained
in Section {\rm\ref{sec:2}}.
\end{remark}
\begin{remark}
In the case $\eta_0=1/6$ $\theta$-parameters of solutions
constructed in this subsection are as follows: $\theta_0=1/3$,
$\theta_1=1/3$, $\theta_t=1/2$, $\theta_\infty=-1/2$. Therefore,
by interchanging points $1$ and $t$ in Eq.~{\rm(\ref{eq:4})}, we
can construct the solutions of $P_6$ for the same case:
$\theta_0=1/3$, $\theta_1=1/2$, $\theta_t=1/3$,
$\theta_\infty=-1/2$ as in the previous subsection. Consider this
in more detail to check that these solutions are different from
the ones constructed previously.

Transformation reads:
$$
\hat t=1/t,\qquad\hat\lambda=\lambda/t,\qquad
\hat\Psi(\hat\lambda)=\hat t^{-\frac{\theta_\infty}2\sigma_3}
\Psi(\hat\lambda/\hat t),
$$
$$
\hat\theta_0=\theta_0=2\eta_0,\quad\hat\theta_1=\theta_t=1/2,\quad
\hat\theta_t=\theta_1=2\eta_0,\quad\hat\theta_\infty=\theta_\infty=-1/2.
$$
This transformation generates the following solutions of $P_6$ and
related functions $\sigma$ and $\tau$, with $\theta$-parameters
changed to $\hat\theta$ parameters and $t$ to $\hat t=1/t$:
\begin{eqnarray*}
\hat y_{12}&=&
\frac{
(2s^2-1)(4\eta_0^2(2s^2+2s+1)^2-s^2(2s+1)^2)}
{s(2s+1)((4\eta_0^2-1)(2s^2+2s+1)^2+3s(s+1)(2s+1))},\\
\hat y_{21}&=&
\frac{2s^2-1}{s(2s+1)}=1-\sqrt{1-\hat t},
\end{eqnarray*}
\begin{eqnarray*}
\hat\sigma&=&
\frac{(s+1)^2-16\eta_0^2s(2s+1)(2s^2+3s+2)}{16s^2(2s+1)^2},\\
\hat\tau&=&\hat t^{-\frac1{16}}
\left(\frac{(2s^2-1)^3}{(s+1)^2(2s^2+2s+1)}\right)^{\eta_0^2},
\quad
\mathrm{where}\quad
\hat t=\frac{(2s^2-1)(2s^2+2s+1)}{s^2(2s+1)^2}.
\end{eqnarray*}
The parameter $s$ can be excluded from the above formulae via the
quadratic equation, $s(2s+1)\sqrt{1-\hat t}=(s+1)$.
\end{remark}
\subsubsection
{$RS_4^2(2+1+1|2+2|2+1+1)$\label{sub:4.1.3}}
$R$-Transformation reads:
$$
\mu=\frac{\rho\lambda(\lambda-1)(\lambda-a)^2}{(\lambda-t)(\lambda-b)^2}
\qquad\mathrm{and}\qquad
\mu-1=\frac{\rho(\la-c_1)^2(\la-c_2)^2}{(\la-t)(\la-b)^2},
$$
where
\begin{eqnarray*}
\rho&=&
\frac{(s-1)^2(-s^2+c_1s^2+c_1)}
{(s+1)^2(2c_1^2s^2-3c_1s^2+s^2-4sc_1^2+4c_1s+2c_1^2-c_1)},\\
a&=&
\frac{s(c_1s^3-s^3-2c_1^2s^2+2c_1s^2+4sc_1^2-3c_1s-2c_1^2)}
{(s-1)^2(-s^2+c_1s^2+c_1)},\quad
b=\frac{s^2}{s^2-1},\\
c_2&=&\frac{s^2(c_1s^2-2c_1s+2s+c_1-s^2)}{
2c_1s^2+2s^3-s^4-s^2-2c_1s+c_1s^4-2c_1s^3+c_1},\\
\end{eqnarray*}
and
$$
t=\frac{(c_1s^2-2c_1s+2s+c_1-s^2)^2c_1^2}
{(2c_1^2s^2-3c_1s^2+s^2-4sc_1^2+4c_1s+2c_1^2-c_1)(-s^2+c_1s^2+c_1)},
$$
with the parameters $s$ and $c_1\in\mathbb{C}$. For the choice of
$\eta$-parameters: $\eta_0=1/2$, $\eta_1=1/2$, and
$\eta_\infty=-1/2$, $RS$-transformation can be written as follows
$$
\Psi(\lambda)=
\left( J_a \sqrt{\frac{\la-a}{\la-c_2}}
+J_{c_2} \sqrt{\frac{\la-c_2}{\la-a}}\right)
\left( J_b \sqrt{\frac{\la-b}{\la-c_1}}
+J_{c_1} \sqrt{\frac{\la-c_1}{\la-b}}\right)
\Phi(\mu),
$$
where
\begin{eqnarray*}
J_{c_1}&=&\bm 0 & -1 \\ 0 & 1 \em, \qquad
J_b=J_{c_1}^*,\qquad J_a=1-J_{c_2},\\
J_{c_2}&=&\frac1{4s(c_1s-s-c_1)}\left(\begin{array}{cc}
(c_1s^2-s^2-c_1)(s+1)&c_1(s-1)^3+s^2(3-s)\\
-(c_1s^2-s^2-c_1)(s+1)&-c_1(s-1)^3-s^2(3-s)
\end{array}\right).
\end{eqnarray*}
The function $\Psi$ solves Eq.~(\ref{eq:4}) with the following parameters:
$$
\theta_0=1/2,\quad \theta_1=1/2, \quad
\theta_t=1/2,\quad \theta_\infty=-1/2.
$$
Solutions of $P_6$ and related functions are:
\begin{eqnarray*}
y_{12}\!&=&\!
\frac{c_1(c_1(s-1)^2+2s-s^2)
(2c_1^2(s-1)^2+c_1+4c_1s-c_1s^2-s^2)}
{3(2c_1^2(s-1)^2-3c_1s^2+s^2+4c_1s-c_1)(-s^2+c_1s^2+c_1)}=
t-\frac13\sqrt{t(t-1)},\\
y_{21}\!&=&\!
\frac{c_1(c_1(s-1)^2+2s-s^2)}{-s^2+c_1s^2+c_1}=
t+\sqrt{t(t-1)},
\end{eqnarray*}
\begin{eqnarray*}
\sigma&=&-\frac{-s^2+c_1s^2+c_1}{16(2c_1^2(s-1)^2-3c_1s^2+s^2+4c_1s-c_1)}=
-\frac1{16}\,\frac{t-\sqrt{t(t-1)}}{t+\sqrt{t(t-1)}},\\
\tau&=&
\frac{(2c_1^2(s-1)^2-3c_1s^2+s^2+4c_1s-c_1)^{\frac 14}}
{\left(c_1(c_1-1)(c_1(s-1)^2-s^2)
(c_1(s-1)^2+2s-s^2)\right)^{\frac 18}}=
\left(\frac{(\sqrt{t}+\sqrt{t-1})^2}{\sqrt{t}\,\sqrt{t-1}}\right)^{\frac18}.
\end{eqnarray*}
The branches of the square roots are chosen such that,
$\sqrt{t}\sqrt{t-1}=\sqrt{t(t-1)}$.
\begin{remark}
 \label{rem:6}
This $RS$-transformation is a combination of two quadratic
transformations. Corresponding solutions of $P_6$ are the special
case of those obtained in Section {\rm\ref{sec:2}} for
$\eta_0=-\eta_\infty=1/2$.
\end{remark}
\subsubsection
{$RS_4^2(1+1+1+1|2+2|4)$}
$R$-Transformation reads:
$$
\mu=\frac{\rho\lambda(\lambda-1)(\lambda-t)}{(\lambda-b)^4}
\qquad\mathrm{and}\qquad
\mu-1=-\frac{(\la-c_1)^2(\la-c_2)^2}{(\la-b)^4},
$$
where
$$
\rho=\frac{(s^2-2)^2}{(s+1)^2},\quad
b=\frac{s(s+2)}{2(s+1)},\quad
c_1=\frac{s^2}2,\quad
c_2=\frac{(s+2)^2}{2(s+1)^2},
$$
and
$$
t=\frac{s^2(s+2)^2}{4(s+1)^2}.
$$
With this $R$-transformation one can associate two different seed
$RS$-transformations: {\bf\thesubsubsection.A} The first
$RS$-transformation can be defined by making the following choice
of the $\eta$-parameters: $\eta_0\in\mathbb{C}$ is arbitrary,
$\eta_1=1/2$, $\eta_\infty=-1/4$. Corresponding
$RS$-transformation can be written as follows,
$$
\Psi(\lambda)=\left( J_t^* \sqrt{\frac{\la-c_1}{\la-t}}
+J_{c_1}^* \sqrt{\frac{\la-t}{\la-c_1}}\right)
\left( J_b \sqrt{\frac{\la-b}{\la-c_2}}
+J_{c_2} \sqrt{\frac{\la-c_2}{\la-b}}\right)
T^{-1}\Phi(\mu),
$$
where
$$
T=\bm
\frac{(4\eta_0+3)(4\eta_0-1)}{16\eta_0} &1 \\
\frac{(4\eta_0-3)(4\eta_0+1)}{16\eta_0} & 1\em,
$$
and
\begin{eqnarray*}
J_b&=& \bm\frac{4\eta_0-1}{8\eta_0}&
-\frac2{4\eta_0-3}\\
-\frac{(4\eta_0-3)(4\eta_0-1)(4\eta_0+1)}{128\eta_0^2}&
\frac{4\eta_0+1}{8\eta_0}\em,\qquad
J_{c_2}=J_b^*,\\
J_{c_1}&=&
\bm -\frac{4\eta_0+1}{4s\eta_0} &
\frac{4(4\eta_0s+4\eta_0+1)}{s(4\eta_0-3)(4\eta_0-1)} \\
-\frac{(4\eta_0-3)(4\eta_0-1)(4\eta_0+1)}{64s\eta_0^2} &
 \frac{4\eta_0s+4\eta_0+1}{4s\eta_0}\em,\qquad
J_t=1-J_{c_1}.
\end{eqnarray*}
The function $\Psi$ solves Eq.~(\ref{eq:4}) with the parameters:
$$
\theta_0=\eta_0,\quad \theta_1=\eta_0, \quad
\theta_t=\eta_0+1,\quad \theta_\infty=\eta_0.
$$
Corresponding solutions of $P_6$ and related functions are as follows:
\begin{eqnarray*}
y_{12}&=&-\frac{s(s+2)(-s^2+8\eta_0s+8\eta_0+2)}
{2(s+1)(-3s^2+8\eta_0s-8s-2+8\eta_0)},\\
y_{21}&=&-\frac{s(s+2)(s^2+4\eta_0s^2+8\eta_0s-2)}
{2(s+1)(3s^2+4\eta_0s^2+8s+8\eta_0s+2)},\\
\sigma&=&
\frac{32(s+1)^2\eta_0^2-16(s+1)(s^2+s-1)\eta_0-s^4-8s^3-12s^2+4}
{64(s+1)^2},\\
\tau&=&\frac{(s+1)^{\frac 18}(s^2+4s+2)^{\frac18+\frac34\eta_0+\frac12\eta_0^2}}
{[s(s+2)]^{\frac 18+\frac12\eta_0+\eta_0^2}(s^2-2)^{\frac 18+\frac14\eta_0-\frac12\eta_0^2}}\\
&=&\left(\frac{\sqrt{t}+1}{2\sqrt{t}(\sqrt{t}-1)}\right)^{\frac18}
\left(\frac{(\sqrt{t}+1)^3}{t(\sqrt{t}-1)}\right)^{\frac{\eta_0}4}
\left(1-\frac1t\right)^{\frac{\eta_0^2}2}.
\end{eqnarray*}
\begin{remark}
We present here also an equivalent $RS$-transformation,
$$
\Psi(\lambda)=\left( J_+^* \sqrt{\la-c_1}
+J_{+,\,c_1}^* \frac1{\sqrt{\la-t}}\right)
\left( J_b \sqrt{\frac{\la-b}{\la-c_2}}
+J_{c_2} \sqrt{\frac{\la-c_2}{\la-b}}\right)
T^{-1}\Phi(\mu),
$$
where the functions, $t=t(s)$ and $\lambda=\lambda(\mu)$,
matrices, $\Phi(\mu)$, $T$, $J_b$, and $J_{c_2}$, are the same as above, and
$$
J_+=\!\left(\!\begin{array}{cc}
1&0\\
0&0
\end{array}\!\right)\!,\quad
J_{+,\,c_1}=\!\left(\!\!\begin{array}{cc}
\frac{(s^2-2)(4\eta_0+1)(4\eta_0s^2+4\eta_0s+3s^2+4s+2)}{64\eta_0(\eta_0+1)(s+1)^2}&
\frac{16\eta_0}{(4\eta_0-3)(4\eta_0-1)}\\
\frac{(4\eta_0-3)(16\eta_0^2-1)(s^2-2)(4\eta_0s^2+4\eta_0s+3s^2+4s+2)}
{1024\eta_0^2(\eta_0+1)(s+1)^2}&1
\end{array}\!\!\right)\!.
$$
This function $\Psi(\lambda)$ solves Eq.~{\rm(\ref{eq:4})} with the parameters:
$$
\theta_0=\eta_0,\quad \theta_1=\eta_0, \quad
\theta_t=\eta_0,\quad \theta_\infty=\eta_0+1,
$$
and generates the following solutions of $P_6$ and related functions:
\begin{eqnarray*}
y_{12}&=&-\frac{s(s+2)}
{2(s+1)}=-\sqrt{t},\\
y_{21}&=&-\frac{s(s+2)(32s(s+1)(s+2)\eta_0^2+4(s^2+4s+2)^2\eta_0+(s^2-2)^2)}
{2(s+1)(32s(s+1)(s+2)\eta_0^2+4(s^2+4s+2)^2(3\eta_0+2)+(s^2-2)^2)},\\
\sigma&=&
\frac{32(s+1)^2\eta_0^2+16(s+1)(s^2+3s+1)\eta_0-s^4+12s^2+16s+4}
{64(s+1)^2},\\
\tau&=&\frac{(s+1)^{\frac 18}(s^2-2)^{\frac18+\frac34\eta_0+\frac12\eta_0^2}}
{[s(s+2)]^{\frac 18+\frac12\eta_0+\eta_0^2}(s^2+4s+2)^{\frac 18+\frac14\eta_0-
\frac12\eta_0^2}}\\
&=&\left(\frac{\sqrt{t}-1}{2\sqrt{t}(\sqrt{t}+1)}\right)^{\frac18}
\left(\frac{(\sqrt{t}-1)^3}{t(\sqrt{t}+1)}\right)^{\frac{\eta_0}4}
\left(1-\frac1t\right)^{\frac{\eta_0^2}2}.
\end{eqnarray*}
\end{remark}
{\bf\thesubsubsection.B} The second $RS$-transformation is defined
by the following choice of the $\eta$-parameters:
$\eta_0\in\mathbb{C}$ is arbitrary, $\eta_1=-\eta_\infty=1/2$.
Corresponding $RS$-transformation reads,
$$
\Psi(\lambda)=
\left( J_b \sqrt{\frac{\la-b}{\la-c_1}}
+J_{c_1} \sqrt{\frac{\la-c_1}{\la-b}}\right)
\left( J_b \sqrt{\frac{\la-b}{\la-c_2}}
+J_{c_2} \sqrt{\frac{\la-c_2}{\la-b}}\right)T^{-1}\Phi(\mu).
$$
In this case the residue matrices in Eq.~(\ref{eq:4}) are as follows:
$$
A_0=\frac{\eta_0}2\si_3,\quad
A_1=-\frac{\eta_0}2\si_3,\quad
A_t=-\frac{\eta_0}2\si_3,
$$
therefore
$$
\theta_0=\eta_0,\quad \theta_1=\eta_0, \quad
\theta_t=\eta_0,\quad \theta_\infty=\eta_0.
$$
In this case solutions of $P_6$ are not defined, the functions
$\sigma$ and $\tau$ are very simple:
$$
\sigma=
\frac 12\eta_0^2,\qquad\qquad
\tau=
\frac{(-2+s^2)^{\frac{\eta_0^2}2}(s^2+4s+2)^{\frac{\eta_0^2}2}}
{s^{\eta_0^2}(s+2)^{\eta_0^2}}=\left(\frac{t-1}t\right)^{\frac{\eta_0^2}2}.
$$
\subsubsection
{$RS_4^2(1+1+1+1|2+2|2+2)$}
$R$-Transformation reads:
$$
\mu=\frac{\rho\lambda(\lambda-1)(\lambda-t)}
{(\lambda-b_1)^2(\la-b_2)^2}
\qquad\mathrm{and}\qquad
\mu-1=-\frac{(\la-c_1)^2(\la-c_2)^2}{(\la-b_1)^2(\la-b_2)^2},
$$
where
$$
\rho=\frac{4(c_1^2+b_2^2-c_1-b_2)}{c_1-b_2},\quad
b_1 = \frac{c_1(c_1+b_2-2)}{b_2-c_1},\quad
c_2=-\frac{b_2(c_1+b_2-2)}{b_2-c_1},
$$
$$
t=\frac{(-2+c_1+b_2)b_2c_1}
{c_1^2+b_2^2-c_1-b_2},
$$
and $c_1$ and $b_2$ are parameters. We present below a seed
$RS$-transformation corresponding to the following choice of the
$\eta$-parameters: $\eta_0\in\mathbb{C}$ is arbitrary,
$\eta_1=-\eta_\infty=1/2$;
$$
\Psi(\lambda)=
\left( J_{b} \sqrt{\frac{\la-b_1}{\la-c_1}}
+J_{c} \sqrt{\frac{\la-c_1}{\la-b_1}}\right)
\left( J_{b} \sqrt{\frac{\la-b_2}{\la-c_2}}
+J_{c} \sqrt{\frac{\la-c_2}{\la-b_2}}\right)
T^{-1}\Phi(\mu).
$$
Here
$$
T=\bm \frac 12\eta_0+\frac 12 &1\\
\frac 12\eta_0-\frac 12 & 1\em,\quad
J_{b}=\frac 1{2}\bm
1 & -\frac 2{\eta_0-1} \\
-\frac 12\eta_0+\frac 12 & 1\em,\quad
J_{c}=J_{b}^*,
$$
Parameters:
$$
\theta_0=\eta_0,\quad \theta_1=\eta_0, \quad
\theta_t=\eta_0,\quad \theta_\infty=\eta_0.
$$
In this case all matrices $A_p$, $p=0,1,t$ are diagonal:
$$
A_0=A_1=-A_t=-\frac{\eta_0}2\sigma_3,
$$
so there is no solution to $y$. The functions $\sigma$ and $\tau$
are very simple:
$$
\sigma=
-\frac{\eta_0^2}2
\frac{(b_2+c_1-1)(2b_2c_1-c_1-b_2)}
{c_1^2+b_2^2-c_1-b_2}=\frac{\eta_0^2}2\left(1-2t\right).
$$
$$
\tau=\left(\frac{(c_1^2+b_2^2-c_1-b_2)^2}
{b_2c_1(b_2-1)(c_1-1)(b_2+c_1)(-2+c_1+b_2)}\right)^{\frac{\eta_0^2}2}
=\left[t(t-1)\right]^{-\frac{\eta_0^2}2}.
$$
\subsubsection
{$RS_4^2(2+1+1|4|2+1+1)$}
$R$-Transformation reads:
$$\mu=
\frac{\rho\lambda
(\lambda-1)(\lambda-a)^2}{(\lambda-t)(\lambda-b)^2}
\qquad\mathrm{and}\qquad
\mu-1=\frac{\rho(\lambda-c)^4}{(\lambda-t)(\lambda-b)^2},
$$
where
\begin{eqnarray*}
\rho&=&\frac{(s^2-4s+2)(s^2-2s+2)^2}{(s^2-2)(3s^2-4s+2)^2},\quad
a=-\frac{s^4(3s^2-8s+6)}{(s^2-4s+2)(s^2-2s+2)^2},\\
b&=&-\frac{s^4}{(s^2-4s+2)(3s^2-4s+2)},\quad
c=\frac{s^3(s-2)}{s^4-6s^3+12s^2-12s+4},
\end{eqnarray*}
$$
t=\frac{s^4(s-2)^4}{(s^2-2s+2)^2(s^2-4s+2)(s^2-2)},
$$
with arbitrary $s\in\mathbb{C}$.There are two different seed
$RS$-transformations which can be associated with this
$R$-transformation:\\
{\bf\thesubsubsection.A} The first transformation is defined by
the following choice of the $\eta$-parameters:
$$
\eta_0=\frac12,\quad\eta_1=\frac12,\quad\mathrm{and}\quad
\eta_\infty=-\frac12.
$$
$RS$-transformation reads,
$$
\Psi(\lambda)=\left( 1-\frac{J_c}{\la-c}\right)\left( J_b
\sqrt{\frac{\la-b}{\la-a}}+J_{a}\sqrt{\frac{\la-a}{\la-b}}\right)
\Phi(\mu),
$$
where
$$
J_b=\bm 1 & -1 \\ 0 & 0 \em, \quad J_a=J_b^*,
$$
and
$$
J_c= j
\bm
1 & -\frac{s^2-2}{3s^2-4s+2}\\
\frac{3s^2-4s+2}{s^2-2} &-1\em,\quad
j=
-\frac{s^2(s-1)^2(s^2-2)}
{(s^2-4s+2)(s^2-2s+2)^2}.
$$
Parameters:
$$
\theta_0=\frac12,\quad\theta_1=\frac12,\quad
\theta_t=\frac12,\quad\mathrm{and}\quad\theta_\infty=-\frac12.
$$
Corresponding solutions of $P_6$ and functions $\sigma$ and $\tau$
are as follows:
\begin{eqnarray*}
y_{12}&=&
\frac{s^2(s-2)^2(3s^4-12s^3+16s^2-8s+4)}
{3(s^2-2s+2)^2(s^2-4s+2)(s^2-2)}=
t+\frac13\sqrt{t(t-1)},\\
y_{21}&=&
\frac{s^2(s-2)^2}{(s^2-2s+2)^2}=
t-\sqrt{t(t-1)},\\
\sigma&=&
-\frac{(s^2-2s+2)^2}{16(s^2-2)(s^2-4s+2)}=
-\frac1{16}\,\frac{t+\sqrt{t(t-1)}}{t-\sqrt{t(t-1)}},\\
\tau&=&
\frac{(s^2-2)^{\frac 14}(s^2-4s+2)^{\frac 14}}
{s^{\frac 14}(s-1)^{\frac 14}(s-2)^{\frac 14}}=
\frac{(2\sqrt{t}-2\sqrt{t-1})^{\frac14}}{[t(t-1)]^{\frac1{16}}}.
\end{eqnarray*}
\begin{remark}
Note that solutions constructed in this subsection coincide with
the ones obtained in Subsection {\rm\ref{sub:4.1.3}}, see also
Remark {\rm\ref{rem:6}}.
\end{remark}
{\bf\thesubsubsection.B}
We define another seed $RS$-transformation by the following choice
of the $\eta$-parameters:
$$
\eta_0=\frac12,\quad\eta_1=\frac14,\quad\mathrm{and}\quad
\eta_\infty=-\frac12.
$$
$RS$-transformation reads,
$$
\Psi(\lambda)=\left( J_c^* \sqrt{\frac{\la-t}{\la-c}}
+J_t^* \sqrt{\frac{\la-c}{\la-t}}\right)
\left( J_b \sqrt{\frac{\la-b}{\la-a}}
+J_{a} \sqrt{\frac{\la-a}{\la-b}}\right) \Phi(\mu).
$$
Here
$$
J_b=\bm 1 & -1 \\ 0 & 0 \em,\quad
J_a=J_b^*,\qquad
J_c=\bm 0 & -\frac{s^2-4s+6}{3s^2-4s+2} \\
0 & 1\em,\quad
J_t= 1- J_c.
$$
Parameters:
$$
\theta_0=\frac12,\quad \theta_1=\frac12, \quad
\theta_t=\frac12,\quad \theta_\infty=-\frac12.
$$
Solutions of $P_6$ and functions $\sigma$ and $\tau$ are as follows:
\begin{eqnarray*}
y_{21}&=&
\frac{s^3(s-2)(3s^2-8s+6)}{(s^2-2s+2)(3s^2-4s+2)(s^2-2)},\\
y_{12}&=&
y_{21}\frac{(3s^2-4s+2)(7s^6-44s^5+106s^4-112s^3+20s^2+80s-72)}
{3(s^2-4s+2)(7s^6-36s^5+86s^4-112s^3+100s^2-80s+40)},\\
\sigma&=&
\frac{s^6-24s^5+122s^4-288s^3+364s^2-224s+56}
{64(s^2-2s+2)^2(s^2-4s+2)},\\
\tau&=&
\frac{(s-2)^{\frac 1{16}}(s^2-4s+2)^{\frac 14}(s^2-2s+2)^{\frac 18}}
{s^{\frac 7{16}}(s-1)^{\frac 7{16}}}.
\end{eqnarray*}
\begin{remark}
In terms of $t$ parameter $s$ reads,
$$
s=1+\sqrt[4]{\frac t{t-1}}+\sqrt{1+\sqrt{\frac t{t-1}}}.
$$
\end{remark}
\subsubsection
{$RS_4^2(2+1+1|2+2|3+1)$} $R$-Transformation reads:
\begin{equation}
 \label{417}
\mu=\frac{\rho\lambda(\lambda-1)(\lambda-t)^2}{(\lambda-b)^3}
\qquad\mathrm{and}\qquad
\mu-1=\frac{\rho(\lambda-c_1)^2(\lambda-c_2)^2}{(\lambda-b)^3}.
\end{equation}
where
\begin{eqnarray*}
\rho&=&\frac{s^3(s^2+1)^3}{(s^4+1)^3},\qquad\qquad\quad\;\,
b=\frac{(s+1)^4(s^2-s+1)^2}{4s(s^2+1)(s^4+1)},\\
c_1&=&\frac{(s+1)^2(s^2-s+1)^2}{4s^3},\quad
c_2=\frac{(s+1)^4(s^2-s+1)}{2(s^2+1)^3},
\end{eqnarray*}
and
$$
t=-\frac{(s+1)^4(s^2-s+1)^2(s^4-2s^3-2s+1)}{4s^3(s^2+1)^3}.
$$
With this $R$-transformation one can associate two different seed
$RS$-transformations:
{\bf\thesubsubsection.A}
The first seed $RS$-transformation can be associated with the
following choice of the $\eta$-parameters:
$$
\eta_0\in\mathbb{C},\quad\eta_1=\frac12,\quad\eta_\infty=-\frac13.
$$
Corresponding $RS$-transformation reads,
$$
\Psi(\lambda)=\left( J_{c_1}^* \sqrt{\frac{\la-1}{\la-c_1}}
+J_1^* \sqrt{\frac{\la-c_1}{\la-1}}\right)
\left( J_b \sqrt{\frac{\la-b}{\la-c_2}}
+J_{c_2} \sqrt{\frac{\la-c_2}{\la-b}}\right)\Phi(\mu).
$$
Here
$$
J_b = \bm 1 & -\frac{6\eta_0-1}{6\eta_0-5}\\
0 & 0\em,\quad J_{c_2}=J_b,\qquad
J_{c_1}=
\bm j & \frac{(1-j)(6\eta_0-1)}{6\eta_0-5}\\
\frac{(6\eta_0-5)j}{6\eta_0-1}& 1-j\em,\quad J_1=1-J_{c_1},
$$
$$
j=-\frac{s(s^4+1)(6\eta_0-1)}{2(s^2+s+1)(s^2+1)^2}.
$$
Parameters:
$$
\theta_0=\eta_0,\quad \theta_1=\eta_0-1, \quad
\theta_t=2\eta_0,\quad \theta_\infty=\eta_\infty.
$$
Solutions:
\begin{eqnarray*}
y_{12}\!\!\!\!&=&\!\!-
\frac{(s+1)^2(s^2-s+1)(s^4-2s^3-2s+1)}
{2s^2(s^2+1)^2}\times\\
&&\!\!\frac{(6s(s^2+1)(s^4+1)\eta_0+(s^2+1)^4+4s^4)
(6(s^4+1)^2\eta_0-(s^2+1)^4-4s^4)}
{(36(s^4+1)^4\eta_0^2+48s(s^2+1)(s-1)^2(s^4+1)^2(s^2+s+1)\eta_0+Q(s))},\\
Q(s)\!\!\!\!&=&\!\!-9s^{16}-8s^{15}+40s^{14}-40s^{13}+28s^{12}-120s^{11}
+56s^{10}-152s^9+\\
&&\!\!+10s^8-152s^7+56s^6-120s^5+28s^4-40s^3+40s^2-8s-9,\\
y_{21}\!\!\!\!&=&\!\!-
\frac{(s+1)^2(s^4+1)(s^2-s+1)(s^4-2s^3-2s+1)(6s(s^2+1)\eta_0+s^4+1)}
{2s^2(s^2+1)^2(6(s^4+1)^2\eta_0+s^8+4s^7-4s^6+4s^5-6s^4
+4s^3-4s^2+4s+1)},\\
\sigma\!\!&=&\!\!\!\!-\frac{3(s^4+1)}{4s(s^2+1)}\eta_0^2
-\frac{s^8-s^7-s^6-s^5-s^3-s^2-s+1}{4s^2(s^2+1)^2}+\\
&&\!\!\!\!\frac{4s^{12}+6s^{11}-15s^{10}-18s^8-6s^7-30s^6-6s^5-18s^4
-15s^2+6s+4}{144s^3(s^2+1)^3}\\
\tau\!\!&=&\!\!\frac{s^{\frac13}(s^2+1)^{\frac13}
(s^4+2s^3+2s+1)^{\frac19-\frac54\eta_0+\frac32\eta_0^2}
\left((s^2+1)^2-s^2\right)^{-\frac{11}{72}+\frac12\eta_0-\frac32\eta_0^2}}
{(s^4-2s^3-2s+1)^{\frac5{36}-\frac14\eta_0-\frac32\eta_0^2}
(s^2-1)^{\frac{11}{36}-\eta_0+3\eta_0^2}}.
\end{eqnarray*}
{\bf\thesubsubsection.B}
To associate another seed $RS$-transformation with
$R$-transformation (\ref{417}) we exchange notation:
$t\longleftrightarrow b$, so that now:
$$
t=\frac{(s+1)^4(s^2-s+1)^2}{4s(s^2+1)(s^4+1)},\quad
b=-\frac{(s+1)^4(s^2-s+1)^2(s^4-2s^3-2s+1)}{4s^3(s^2+1)^3}.
$$
and the other parameters in (\ref{417}) remain unchanged.
$RS$-transformation can be defined by the following choice of
parameters:
$$
\eta_0=\eta_1=\frac 12,
$$
and arbitrary $\eta_\infty\in\mathbb{C}$. It reads,
$$
\Psi(\lambda)=\left( J_{c_1}^* \sqrt{\frac{\la-1}{\la-c_1}}
+J_1^* \sqrt{\frac{\la-c_1}{\la-1}}\right)
\left( J_a \sqrt{\frac{\la-a}{\la-c_2}}
+J_{c_2} \sqrt{\frac{\la-c_2}{\la-a}}\right)\Phi(\mu).
$$
Here
$$
J_a=
\frac 12 \bm 1 &1 \\ 1 & 1\em, \quad J_{c_2}=J_a^*,\quad
J_{c_1}=J_{c_2}, \quad J_1=1-J_{c_1}.
$$
Parameters:
$$
\theta_0=\frac 12,\quad \theta_1=\frac 12, \quad
\theta_t=3\eta_\infty,\quad \theta_\infty=\eta_\infty.
$$
Corresponding solutions of $P_6$ and functions $\sigma$ and $\tau$
are as follows:
\begin{eqnarray*}
y_{12}&=&
-\frac{(s+1)^2(s^2-s+1)
(2(s^4-2s^3-2s+1)\eta_\infty+s^4+s^3+s+1)}
{4(\eta_\infty-1)s(s^2+1)(s^4+1)},\\
y_{21}&=&-\frac{(s+1)^2(s^2-s+1)
(2(s^4-2s^3-2s+1)\eta_\infty-s^4-s^3-s^2-s-1)}
{4(\eta_\infty+1)s(s^2+1)(s^4+1)},\\
\sigma&=&
-\frac{\eta_\infty^2(s^4+2s^3+4s^2+2s+1)(s^4-2s^3+4s^2-2s+1)}
{4s(s^2+1)(s^4+1)},\\
\tau&=&
\left(\frac{s(s^2+1)(s^4+1)^4}
{(s^2-1)^5(s^2+s+1)^{\frac 52}(s^2-s+1)^{\frac 52}}
\right)^{\eta_\infty^2}.
\end{eqnarray*}
\begin{subsection}
{$RS$-transformations with fixed $t$}
One proves that the triple
$(3+1|2+2|2+2)$ does not define any $R$-transformation; therefore
only six triples of those seven mentioned in the beginning of
section \ref{sec:4} define $RS$-transformations with fixed $t$.
\end{subsection}
\subsubsection
{$RS_4^2(2+1+1|2+2|4)$}
{\bf\thesubsubsection.A} $R$-Transformation reads,
$$
\mu=\frac{\lambda(\lambda-1)(\lambda-a)^2}{(\lambda-b)^4}
\qquad\mathrm{and}\qquad
\mu-1=-\frac{(\lambda-t)^2}{2(\lambda-b)^4},
$$
where
$$
a=2b-\frac12,\quad b=\frac12\pm\frac{\sqrt{2}}4,\quad
t=\frac32b-\frac14.
$$
To construct $RS$-transformation one chooses $\eta$-parameters as
follows: $\eta_0=\frac12$, $\eta_\infty=\frac12$ (or $\frac14$),
and $\eta_1\in\mathbb{C}$ is arbitrary. This allows one to map
Eq.~(\ref{eq:3}) into Eq.~(\ref{eq:4}) with the parameters:
$$
\theta_0=\frac 12,\quad \theta_1=\frac 12,\quad \theta_t=2\eta_1,
\quad \theta_\infty = 2\eta_1,\quad\mathrm{and}\quad
t=\frac12\pm\frac{3\sqrt{2}}8.
$$
It can be presented as a superposition of two
$RS$-transformations of the rank $2$.\\
{\bf\thesubsubsection.B} Another equivalent form of this
$R$-transformation can be written as follows,
$$
\mu=-4\lambda^2(\lambda-1)(\lambda-t)
\qquad\mathrm{and}\qquad
\mu-1=-4(\lambda-\frac 1{\sqrt 2})^2(\lambda+\frac 1{\sqrt 2})^2,
$$
where $t=-1$. One can define $RS$-transformations by making the
following choice of $\eta$-parameters: $\eta_0$ and
$\eta_\infty\in\mathbb{C}$ are arbitrary and $\eta_1=\frac 12$.
This $RS$-transformation maps Eq.~(\ref{eq:3}) into
Eq.~(\ref{eq:4}) with the following parameters:
$$
\theta_0=2\eta_0,\quad \theta_1=\theta_t=\eta_0,\quad
\theta_\infty = 4\eta_\infty,\quad\mathrm{and}\quad
t=-1.
$$
It is also a combination of two $RS$-transformations of the rank
$2$.
\subsubsection
{$RS_4^2(3+1|4|2+1+1)$}
{\bf\thesubsubsection.A}
$$
\mu=\frac{\rho\lambda(\lambda -1)^3}
{(\lambda-t)(\lambda-b)^2}
\qquad\mathrm{and}\qquad
\mu-1=\frac{\rho(\lambda-c)^4}{(\lambda-t)(\lambda-b)^2},
$$
where $\rho=-\frac29\pm\frac{i}{9\sqrt{2}}$ and
$$
t=-\frac{63}{8}\rho-\frac54,\qquad b=\frac14+\frac34\rho,\qquad
c=-\frac54-\frac92\rho.
$$
By making the following choice of the $\eta$-parameters:
$\eta_1=1/4$, $\eta_\infty=1/2$, and arbitrary
$\eta_0\in\mathbb{C}$, one defines $RS$-transformation which
removes apparent singularities, $b$ and $c$. Corresponding
$\theta$-parameters of the resulting Eq.~(\ref{eq:4}) read:
$$
\theta_0=\eta_0,\quad\theta_1=3\eta_0,\quad\theta_t=\frac12,\quad
\theta_\infty=\frac12,\quad\mathrm{and}\quad
t=\frac12\mp\frac{7i\sqrt{2}}{16}.
$$
{\bf\thesubsubsection.B} It is convenient to consider also another
form of the $R$-transformation,
$$
\mu=\frac{\rho
\lambda(\lambda-a)^3}{(\lambda-t)(\lambda-b)^2}
\qquad\mathrm{and}\qquad
\mu-1= \frac{\rho(\lambda-1)^4}{(\lambda-t)(\lambda-b)^2},
$$
where $\rho=\frac1{12}\pm\frac{i}{{12}\sqrt{2}}$ and
$$
t=30\rho-2,\qquad a=32\rho-4,\qquad b=\frac43\rho-\frac13.
$$
Choosing the $\eta$-parameters: $\eta_0=1/3$, $\eta_\infty=1/2$,
and $\eta_1\in\mathbb{C}$ is arbitrary, one defines
$RS$-transformation which removes apparent singularities, $a$ and
$b$. Corresponding $\theta$-parameters of the resulting
Eq.~(\ref{eq:4}) read:
$$
\theta_0=\frac13,\quad\theta_1=4\eta_1,\quad\theta_t=\frac12,\quad
\theta_\infty=\frac12,\quad\mathrm{and}\quad
t=\frac12\pm\frac{5i\sqrt{2}}4.
$$
{\bf\thesubsubsection.C} We consider here one more equivalent form
of the same $R$-transformation,
$$
\mu=\frac{\rho \lambda(\lambda-a)^3}{(\lambda-t)(\lambda-1)^2}
\qquad \mathrm{and}\qquad
\mu-1=\frac{\rho(\lambda-c)^4}{(\lambda-t)(\lambda-1)^2},
$$
where $c=-4\pm i\sqrt{2}$ and
$$
t=\frac{45}2+\frac{11}2c,\qquad\rho=\frac{c-1}{216},\quad a=4c+24.
$$
Putting the $\eta$-parameters: $\eta_0=1/3$ or $2/3$, $\eta_1=1/4$
or $1/2$, and $\eta_\infty\in\mathbb{C}$ is arbitrary, one defines
$RS$-transformation which removes apparent singularities, $a$ and
$c$. There are two non-equivalent $RS$-transformations which
define two different Eqs.~(\ref{eq:4}) with the following
$\theta$-parameters:
$$
\theta_0=\frac13\;\;\mathrm{or}\;\;\frac23,\quad\theta_1=2\eta_\infty,
\quad\theta_t=\eta_\infty,\quad\theta_\infty=\eta_\infty,\quad\mathrm{and}
\quad t=\frac12\pm\frac{11i\sqrt{2}}2.
$$
\subsubsection
{$RS_4^2(3+1|3+1|3+1)$}
$$
\mu=-\frac{\lambda(\lambda-1)^3}{4(\lambda-1/4)^3}
\qquad\mathrm{and}\qquad
\mu-1=-\frac{(\lambda-t)(\lambda+1/2)^3}{4(\lambda-1/4)^3},
$$
where $t=1/2$. $RS$-Transformation is defined by the following
choice of the $\eta$-parameters: arbitrary $\eta_0\in\mathbb{C}$,
$\eta_1=1/3$, and $\eta_\infty=1/3$ or $2/3$.

There are two non-equivalent $RS$-transformations which define
Eq.~(\ref{eq:4}) with the following $\theta$-parameters:
$$
\theta_0=\eta_0,\quad\theta_1=3\eta_0,\quad\theta_t=\frac13,\quad
\theta_\infty=\frac13\;\;\mathrm{or}\;\;\frac23,\quad\mathrm{and}
\quad t=\frac12.
$$
\subsubsection
{$RS_4^2(3+1|3+1|2+2)$}
{\bf\thesubsubsection.A}
$$
\mu=\frac{\rho\lambda(\lambda-a)^3}{(\lambda-t)^2}
\qquad\mathrm{and}\qquad
\mu-1=\frac{\rho(\lambda-1)(\lambda-c)^3}
{(\lambda-t)^2},
$$
where
$$
\rho=\pm\frac32\sqrt{3}, \quad t=\frac12-\frac5{27}\rho,\quad
a=\frac23-\frac4{27}\rho,\quad c=\frac13-\frac4{27}\rho.
$$
$RS$-Transformation is defined by the following choice of the
$\eta$-parameters: $\eta_0=1/3$ or $2/3$, $\eta_1=1/3$, and
$\eta_\infty\in\mathbb{C}$ is arbitrary. There are two
non-equivalent $RS$-transformations which define Eq.~(\ref{eq:4})
with the following $\theta$-parameters:
$$\theta_0=\frac13\;\;\mathrm{or}\;\;\frac23,\quad\theta_1=\frac13,\quad
\theta_t=2\eta_\infty,\quad\theta_\infty=2\eta_\infty\quad\mathrm{and}
\quad t=\frac12\mp\frac{5i\sqrt{3}}{18}.
$$
{\bf\thesubsubsection.B} Another form of this $R$-transformation
reads,
$$
\mu= \frac{\lambda(\lambda-1)^3}
{(\lambda-b_+)^2(\lambda-b_-)^2}
\qquad\mathrm{and}\qquad
\mu-1=-8\frac{(\lambda-t)^3} {(\lambda-b_+)^2(\lambda-b_-)^2},
$$
where $t=-1/8$ and $b_{\pm}=-5/2\pm3\sqrt{3}/2$. Corresponding
$RS$-transformation is defined by taking arbitrary
($\in\mathbb{C}$) parameters $\eta_0$ and $\eta_1$, and putting
$\eta_\infty=1/2$. The $\theta$-parameters of the resulting
Eq.~(\ref{eq:4}) are as follows:
$$
\theta_0=\eta_0,\quad\theta_1=3\eta_0,\quad\theta_t=3\eta_1,
\quad\theta_\infty=\eta_1,\quad\mathrm{and}\quad t=-\frac18.
$$
{\bf\thesubsubsection.C} Consider here one more form of the same
$R$-transformation,
$$
\mu=\frac{\rho\lambda(\lambda-1)^3}{(\lambda-b)^2}
\qquad\mathrm{and}\qquad
\mu-1=\frac{\rho(\lambda-t)(\lambda-c)^3}{(\lambda-b)^2},
$$
where $c=\frac14\pm\frac{\sqrt{3}}4$,
$$
t=3-3c,\quad b=\frac14+c,\quad \rho=-\frac49-\frac{32}9c.
$$
$RS$-Transformation is defined by taking arbitrary
$\eta_0\in\mathbb{C}$, $\eta_1=1/3$, and $\eta_\infty=1/2$.
Corresponding $\theta$-parameters of Eq.~(\ref{eq:4}) read:
$$
\theta_0=\eta_0,\quad\theta_1=3\eta_0,\quad\theta_t=1/3,\quad
\theta_\infty=1,\quad\mathrm{and}\quad
t=\frac94\mp\frac{3\sqrt{3}}4.
$$
Note that in this case singularity of Eq.~(\ref{eq:4}) at
$\lambda=\infty$ is apparent.
\subsubsection
{$RS_4^2(1+1+1+1|4|4)$}
$R$-Transformation reads:
$$
\mu=\frac{\rho\lambda(\lambda-1)(\lambda-t)}{(\lambda-b)^4}
\qquad\mathrm{and}\qquad
\mu-1=-\frac{(\lambda-c)^4}{(\lambda-b)^4}.
$$
As a result of fractional linear transformation of the complex
$\lambda$ - plane interchanging $0$, $1$, and $\infty$, there are
three sets of possible values for the parameters:
\begin{enumerate}
\item
$c=\pm \imath, \quad b= \mp \imath,\quad \rho = \pm 8 \imath,\quad
t=-1$;
\item
$c=1\pm \imath,\quad b=1\mp \imath,\quad \rho =\pm 8 \imath,\quad
t=2$;
\item
$c=\frac{1\pm \imath}2,\quad b=\frac{1\mp \imath}2,\quad \rho=\pm
4\imath, \quad t= \frac12$.
\end{enumerate}
There are two (non-equivalent) $RS$-transformations:\\
{\bf\thesubsubsection.A} The first $RS$-transformation is defined
by taking arbitrary $\eta_0\in\mathrm{C}$ and putting
$\eta_1=\eta_\infty=1/4$. Resulting Eq.~(\ref{eq:4}) has the
following parameters:
$$
\theta_0=\eta_0,\quad\theta_1=\eta_0,\quad\theta_t=\eta_0,
\quad\theta_\infty=\eta_0,\quad\mathrm{and}\quad
t=-1,\;2,\;\mathrm{or}\;1/2.
$$
{\bf\thesubsubsection.B} The second $RS$-transformation is defined
by taking arbitrary $\eta_0\in\mathrm{C}$ and putting $\eta_1=1/4$
and $\eta_\infty=1/2$. The parameters of the resulting
Eq.~(\ref{eq:4}) are as follows:
$$
\theta_0=\eta_0,\quad\theta_1=\eta_0,\quad\theta_t=\eta_0,
\quad\theta_\infty=\eta_0+1,\quad\mathrm{and}\quad
t=-1,\;2,\;\mathrm{or}\;1/2.
$$
\subsubsection
{$RS_4^2(2+2|2+2|2+2)$}
$R$-transformation reads:
$$
\mu=-\frac{\lambda^2(\lambda-1)^2}{(\la-1/2)^2}
\qquad\mathrm{and}\qquad
\mu-1=-\frac{(\la-1/2-i/2)^2(\la-1/2+i/2)^2}{(\lambda-1/2)^2}.
$$
One can define two (non-equivalent) $RS$-transformations:\\
{\bf\thesubsubsection.A} The first $RS$-transformation is defined
by taking arbitrary ($\in\mathrm{C}$) $\eta_0$ and $\eta_\infty$
and putting $\eta_1=1/2$. Resulting Eq.~(\ref{eq:4}) has the
following parameters:
$$
\theta_0=2\eta_0,\quad\theta_1=2\eta_0,\quad\theta_t=2\eta_\infty,
\quad\theta_\infty=2\eta_\infty,\quad\mathrm{and}\quad t=1/2.
$$
{\bf\thesubsubsection.B} The second $RS$-transformation is defined
by taking arbitrary $\eta_0\in\mathrm{C}$ and putting
$\eta_1=\eta_\infty=1/2$. The parameters of the resulting
Eq.~(\ref{eq:4}) are as follows:
$$
\theta_0=2\eta_0,\quad\theta_1=2\eta_0,\quad\theta_t=1,
\quad\theta_\infty=1,\quad\mathrm{and}\quad t=1/2\pm i/2.
$$
In this case singularities of Eq.~(\ref{eq:4}) at $\lambda=t$ and
$\lambda=\infty$ are apparent.

{\bf Acknowledgment} F.~V.~A. is grateful to Lev Kapitanski and
Andrew Bennett for encouragement, his work was supported by NSF
grants \#436--2978 and CMS--9813182. A.~V.~K. was supported by
Alexander von Humboldt-Stiftung and hosted by Universit\"at GH
Paderborn. Some results of this work were presented at Conference
on Differential Equations in the Complex Domain dedicated to the
memory of Professor Raymond G\'erard in Strasbourg, February
20--23, 2001. We are grateful to the scientific and organizing
committees for the invitation and financial support.

\end{document}